\documentclass[prl, showpacs, twocolumn, superscriptaddress]{revtex4}
\usepackage{graphics}

\newcommand{\bra}{\left\langle}
\newcommand{\ket}{\right\rangle}
\newcommand{\ds}{\mathrm{d}s}
\newcommand{\dt}{\mathrm{d}t}

\newcommand{\dx}{\mathrm{d}x}
\newcommand{\dw}{\mathrm{d}\omega}
\newcommand{\pder}[2]{\frac{\partial #1}{\partial  #2}}

\newcommand{\ve}{\varepsilon}
\newcommand{\fp}{f^{\rm p}}

\newcommand{\vs}{v_\mathrm{s}}
\newcommand{\LFP}{\mathcal{L}}
\newcommand{\Ps}{P^\mathrm{st}}

\newcommand{\vtr}[2]{\left( \begin{array}{c} #1 \\ #2 \end{array} \right)}
\newcommand{\mtx}[4]{\left( \begin{array}{cc} #1 & #2 \\ #3 & #4 \end{array} \right)}

\def\Vec#1{\mbox{\boldmath $#1$}}

\begin{document}

\title{Equality connecting energy dissipation with violation of fluctuation-response relation}

\author{Takahiro Harada}
\email[Electronic address: ]{harada@chem.scphys.kyoto-u.ac.jp}
\affiliation{Department of Physics, Graduate School of Science, Kyoto University, Kyoto 606-8502, Japan}

\author{Shin-ichi Sasa}
\email[Electronic address: ]{sasa@jiro.c.u-tokyo.ac.jp}
\affiliation{Department of Pure and Applied Sciences, University of Tokyo, Komaba, Tokyo 153-8902, Japan}

\date{\today}

\begin{abstract}
In systems driven away from equilibrium, the velocity correlation function and the linear response function to a small perturbation force do not satisfy
the fluctuation-response relation (FRR) due to the lack of detailed balance in contrast to equilibrium systems.
In this Letter, an equality between an extent of the FRR violation and the rate of energy dissipation is proved for Langevin systems under non-equilibrium conditions.
This equality enables us to calculate the rate of energy dissipation by quantifying the extent of the FRR violation, which can be measured experimentally.
\end{abstract}

\pacs{05.40.Jc, 05.70.Ln, 87.16.Nn}

\maketitle



Macromolecules and colloidal particles of the order of nanometers to sub-micrometers suspended in an aqueous solution provide an ideal ground to study the foundation of non-equilibrium statistical mechanics. 
Recent advances in experimental techniques for the observation and manipulation of such small systems has generated the possibility 
of directly investigating the non-equilibrium nature of fluctuations in the system.
In particular, such techniques have been designed to verify several universal relations such as the fluctuation theorem \cite{ft, cg, exft}, the Jarzynski equality \cite{jarz, exjarz}, and the Hatano-Sasa identity \cite{hata, exhata}. 


Through the investigation of non-equilibrium systems, it has been recognized that the quantification of the violation of the fluctuation-response relation (FRR) \cite{FDT} 
provides new information for systems driven far from equilibrium \cite{CKP, harada1, HS1}. 
On the other hand,  the rate of energy dissipation is  the most fundamental quantity that characterizes non-equilibrium steady states.
Thus, it is naturally expected that the FRR violation is related to the amount of energy dissipation. 
Toward this end, in this Letter, we present an equality between the rate of energy dissipation and an extent of the FRR violation for a class of non-equilibrium stochastic systems.


In this Letter, for simplicity, we mainly study  a system 
described by the Langevin equation: 
\begin{equation}
\gamma \dot x(t) = F(x(t), t) + \xi(t) + \ve \fp(t),
\label{e.Langevin}
\end{equation}
where $\gamma$ is a friction coefficient, and $\xi(t)$ is zero-mean white Gaussian noise with variance $2 \gamma T$.
In particular, we consider two models of the force $F(x,t)$.
In the first model, which is termed Model A, $F(x, t) \equiv -\delta_{\sigma(t)0} \partial_x U_0 (x) - \delta_{\sigma(t)1} \partial_x U_1(x)$, where $U_i (x)$ for $i = 0, 1$ are periodic potentials with period $\ell$, and $\sigma(t)$ is a Poisson process on  $\{0,1\}$ with a constant transition rate $\alpha$ for the transitions from both $0$ to $1$ and $1$ to $0$.
This model was originally studied as a model of motor proteins, and is termed a {\it flashing ratchet model} \cite{flash1, flash2}.
In the second model, termed Model B, we address a time-independent force $F(x,t) = F(x) \equiv f - \partial_x U(x)$, where $f$ is a constant driving force and $U(x)$ is a periodic potential with period $\ell$. See Ref.~\cite{Risken} for the physical significance of this model. 
The last term in Eq.~(\ref{e.Langevin}) with a sufficiently small $\ve$ represents a ``probe'' force  used to investigate  the linear-response property of the system.
The initial condition is given at $t = t_{\rm init}$ and we consider the limit $t_{\rm init} \to -\infty$ below. 

First, we define several measurable quantities for this system.
Assuming $\vs$ to be the steady-state velocity with $\ve = 0$, the response of the velocity to the probe force is characterized as 
\begin{equation}
\bra \dot x(t) \ket_\ve 
= \vs + \ve \int_{-\infty}^t R(t-s) \fp(s) \ds + O(\ve^2),
\label{e.response}
\end{equation}
where $\bra \cdots \ket_\ve$ denotes the ensemble average in the presence of the probe force with $\ve$. 
The response function $R(t)$ has the causality property $R(t) = 0$ for $t < 0$.
Another important quantity is the time-correlation function of velocity fluctuations in the absence of perturbation, which is defined as 
\begin{equation}
C(t) \equiv \bra \left[ \dot x(t) - \vs \right]  \left[ \dot x(0) - \vs \right] \ket_0.
\label{e.correlation}
\end{equation}


Next, we quantify the energy dissipation in the Langevin description.
According to the definition in Ref.~\cite{Sekimoto}, the rate of energy dissipation $J(t)$ at time $t$ for each trajectory is expressed as 
\begin{equation}
J(t) \dt \equiv \left[ \gamma \dot x(t) - \xi(t) \right] \circ \dx(t) ,
\label{e.heat}
\end{equation}
where  $\circ$ denotes the Stratonovich multiplication \cite{gardiner}.
This definition of the rate of energy dissipation complies with both the conservation of energy and the second law of thermodynamics \cite{2nd}. 


Based on the above, we present a {\it Theorem} that constitutes the main claim of this Letter:
\begin{equation}
\bra J \ket_0 = 
\gamma \left\{ \vs^2 + \int_{-\infty}^\infty 
\left[ \tilde C(\omega) - 2 T \tilde R'(\omega) \right] 
\frac{\dw}{2 \pi} \right\}.
\label{e.equality}
\end{equation}
We express the Fourier transform of an arbitrary function $A(t)$ as $\tilde A(\omega) \equiv \int_{-\infty}^\infty A(t) \exp(i \omega t) \dt$. 
The prime denotes the real part of the function. 
It is widely acknowledged that in equilibrium, i.e., for cases wherein $\alpha = 0$ for Model A and $f=0$ for Model B, the correlation function $C(t)$ is connected to the response 
function $R(t)$ as $C(t) = T R(t)$ for $t > 0$, which is the FRR in model (\ref{e.Langevin}) \cite{FDT, Risken}.
Thus, from $C(t)=C(-t)$ and $R(t)=0$ for $t <0$, we obtain the Fourier transform of this relation as 
\begin{equation}
\tilde C (\omega) = 2T \tilde R'(\omega).
\label{e.FDT}
\end{equation}
Here, we stress that Eq.~(\ref{e.FDT}) does not hold in non-equilibrium steady states; thus, it is observed that the right-hand side of Eq.~(\ref{e.equality}) represents the extent of the FRR violation.


It is noteworthy that the equality (\ref{e.equality}) holds for variety of Langevin systems regardless of the magnitude of external driving as well as the manner in which the system is driven away from equilibrium.
We will provide a detailed explanation of this equality in another paper.
In the present Letter, we demonstrate the validity of Eq.~(\ref{e.equality}) by the numerical verification for Model A and provide a mathematical proof for Model B.

%
%

\begin{figure}[t]
\begin{center}
\scalebox{1.0}{\includegraphics{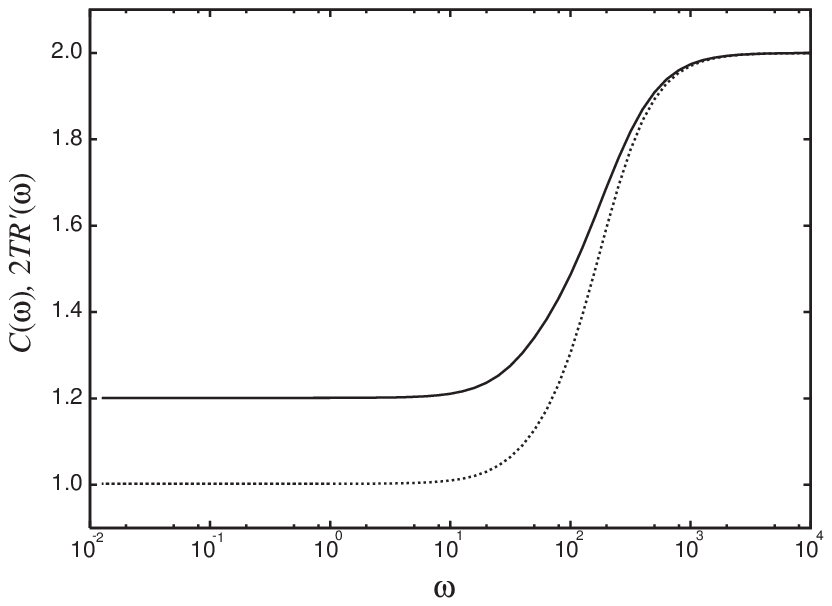}}
\caption{
$\tilde C(\omega)$ and $2T\tilde R'(\omega)$ as functions of  $\omega$ when $D=5$ and $\alpha = 10$.
All the quantities are dimensionless under the normalization as $\gamma = \ell = T = 1$.
The solid and dotted lines represent $\tilde C(\omega)$ and $2T\tilde R'(\omega)$, respectively.
}
\label{f.CR}
\end{center}
\end{figure}
First, we investigate Model A.
Statistical quantities for Model A can be calculated by analyzing the following Fokker-Planck equation that corresponds to the Langevin equation (\ref{e.Langevin}) \cite{flash1, flash2}:
\begin{equation}
\pder{}{t} \vtr{P_0 (x, t)}{P_1 (x, t)} 
= \mtx{\LFP_0 - \alpha}{\alpha}{\alpha}{\LFP_1 - \alpha} 
\vtr{P_0 (x, t)}{P_1 (x, t)},
\label{e.flashFP}
\end{equation}
where $\LFP_i \equiv - \partial_x [ F_i(x) + \ve \fp(t) - T \partial_x ]/\gamma$ and $F_i (x) \equiv - \partial_x U_i (x)$ for $i = 0,1$.
In particular, the potentials were selected as $U_0 (x) = D \cos (2\pi x/\ell)$ and $U_1 (x) = {\rm const.}$, and all the quantities were converted into dimensionless forms by normalizing $\gamma$, $\ell$, and $T$ to unity.
The functions, $\tilde C(\omega)$ and $\tilde R(\omega)$, were calculated accurately by the matrix-continued-fraction method \cite{Risken}.
In Fig.~\ref{f.CR}, $\tilde C(\omega)$ and $2 T \tilde R'(\omega)$ are represented as functions of $\omega$ when $D=5$ and $\alpha = 10$. 
It exemplifies the violation of the FRR in a non-equilibrium steady state.
In this case, it should be noted that $\vs=0$ because the selected potentials are symmetric 
with respect to the reflection of the variable $x$.
We then calculated the right-hand side of Eq.~(\ref{e.equality}), which amounted to 16.891, by integrating the difference between $\tilde C(\omega)$ and $2 T \tilde R'(\omega)$ over the entire frequency domain.
On the other hand, the rate of energy input for Model A has been calculated as \cite{Sekimoto, Parrondo}
\begin{equation}
\bra J_{\rm in} \ket_0 = \alpha \int_0^\ell \left[ U_1(x) - U_0(x) \right] \left[ \Ps_0 (x) - \Ps_1(x) \right] \dx,
\label{e.jflash}
\end{equation}
where $(\Ps_0 (x), \Ps_1(x))$ is the stationary solution of Eq.~(\ref{e.flashFP}).
As a result of the energy balance, the rate of energy input $\bra J_\mathrm{in} \ket_0$ should coincide with the rate of energy dissipation $\bra J \ket_0$ irrespective of the validity of Eq.~(\ref{e.equality}).
Subsequently, if $\bra J_\mathrm{in} \ket_0$ coincides with the right-hand side of Eq.~(\ref{e.equality}), it indicates the validity of Eq.~(\ref{e.equality}).
Using the above parameters, we obtained $\bra J_{\rm in} \ket_0 = 16.891$.
Thus, in this case, Eq.~(\ref{e.equality}) has been confirmed.

\begin{figure}[t]
\begin{center}
\scalebox{1.0}{\includegraphics{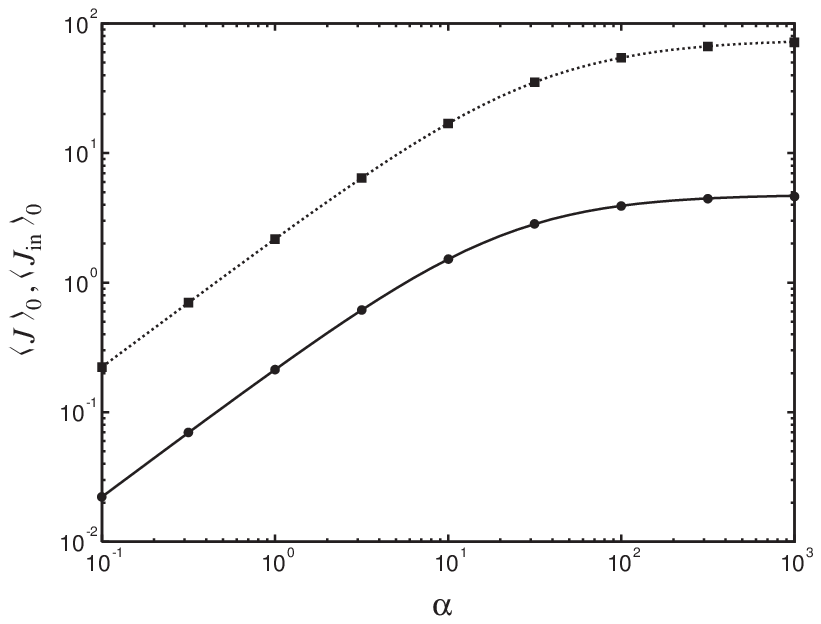}}
\caption{Energy fluxes $\bra J \ket_0$ and $\bra J_{\rm in} \ket_0$ calculated on the basis of Eq.~(\ref{e.equality}) and Eq.~(\ref{e.jflash}), respectively, as functions of the transition rate $\alpha$.
The solid line and the closed circles represent $\bra J \ket_0$ and $\bra J_{\rm in} \ket_0$, respectively, when $D = 1$.
The dotted line and the closed squares represent $\bra J \ket_0$ and $\bra J_{\rm in} \ket_0$, respectively, when $D = 5$.
As in Fig.~\ref{f.CR}, all the quantities are dimensionless.}
\label{f.flash}
\end{center}
\end{figure}
In order to demonstrate that this result is not accidental, both the quantities defined in Eqs.~(\ref{e.equality}) and (\ref{e.jflash}) are represented as functions of $\alpha$ for two values of $D$ in Fig.~\ref{f.flash}.
This figure clearly shows a good agreement between $\bra J \ket_0$ and $\bra J_{\rm in} \ket_0$, independent of the values of the model parameters.
In this manner, Eq.~(\ref{e.equality}) has been numerically verified for Model A.


Next, we provide a mathematical proof of the theorem.
For simplicity, we restrict our argument to  Model B in this proof.
In the case of Model B, it has recently been demonstrated that the response equation (\ref{e.response}) can be derived from the Langevin equation (\ref{e.Langevin}) by transforming the force $F(x(t))$ as \cite{HHS}
\begin{eqnarray}
\lefteqn{
F(x(t)) = \gamma \vs + \int_{-\infty}^t K_0(t-s) \cdot 
\left[ \xi(s) + \ve \fp(s) \right] \ds} \hspace{5mm} \nonumber \\
& & + \int_{-\infty}^t K_\perp (t-s; x(s)) \cdot 
\left[ \xi(s) + \ve \fp(s) \right] \ds,
\label{e.decomp}
\end{eqnarray}
where $\cdot$ denotes the It\^o multiplication \cite{gardiner}.
We define $K_0(t) = 0$ and $K_\perp (t; x(s)) = 0$ for $t < 0$.
This decomposition of force is determined by defining $K_\perp (t; x(s))$ such that it satisfies the property 
\begin{equation}
\bra K_\perp (t; x(s)) \ket_0 = 0
\label{e.Kp0}
\end{equation}
for arbitrary $t > 0$ and $s$. 
By taking the average of Eq.~(\ref{e.Langevin}) after substitution of Eq.~(\ref{e.decomp}),
we obtain Eq.~(\ref{e.response}) by the use of Eq.~(\ref{e.Kp0}). Then, it is established that
\begin{equation}
\gamma \tilde R(\omega) = \tilde K_0 (\omega) + 1.
\label{e.rk0}
\end{equation}
See Ref.~\cite{HHS} for further details.

We now set $\ve = 0$.
Hereafter, in order to distinctly treat the coexistence of two types of multiplication $\circ$ and $\cdot$, we discretize time as $t_n  \equiv n \Delta t$.
Furthermore, we introduce a Wiener process \cite{gardiner}, $W(t)$, in place of the noise, $\xi(t)$.
Accordingly, the following notations are employed: $x_n \equiv x(t_n)$, $\Delta x_n \equiv x_{n+1} - x_n$, and $\Delta W_n \equiv W(t_{n+1})-W(t_n)$.  
By the definition of the Wiener process, the relations, $\bra \Delta W_n \ket_0 = 0$ and $\bra \Delta W_n \Delta W_m \ket_0 = \delta_{nm} \Delta t$, hold.

Combining Eqs.~(\ref{e.Langevin}) and (\ref{e.heat}) and taking into account the definition of the multiplication $\circ$, we obtain 
\begin{equation}
J(t_n) \Delta t = \bar F_n \Delta x_n + O(\Delta t^{3/2}),
\label{e.disheat}
\end{equation}
where $\bar F_n \equiv [F(x_n) + F(x_{n+1})]/2$.
On the other hand, integration of Eq.~(\ref{e.Langevin}) from $t_n$ to $t_{n+1}$ gives
\begin{equation}
\gamma \Delta x_n = \bar F_n \Delta t + \sqrt{2 \gamma T} \Delta W_n + O(\Delta t^2).
\label{e.discrete}
\end{equation}
A straightforward calculation then gives
\begin{eqnarray}
\bra J (t_n) \ket_0 &=& \gamma \vs^2 + \gamma \bra \left( \frac{\Delta x_n}{\Delta t} - \vs \right)^2 \ket_0 - \frac{2T}{\Delta t} \nonumber \\
& & - \sqrt{\frac{2T}{\gamma}}\frac{\bra \bar F_n \Delta W_n \ket_0}{\Delta t} + O(\Delta t^{1/2}).
\label{e.p3}
\end{eqnarray}

In the limit of $\Delta t \to 0$, the second and third terms on the right-hand side of Eq.~(\ref{e.p3}) can be transformed as 
\begin{eqnarray}
\lefteqn{
\lim_{\Delta t \to 0}
\left[ \gamma \bra \left( \frac{\Delta x_n}{\Delta t} - \vs \right)^2 \ket_0 -\frac{2T}{\Delta t} \right] 
} \hspace{20mm} \nonumber \\
&=& \int_{-\infty}^\infty 
\left [ \gamma \tilde C(\omega) -2T \right]\frac{\dw}{2\pi}.
\label{e.cint}
\end{eqnarray}

Next, since the discrete expression of Eq.~(\ref{e.decomp}) becomes
\begin{eqnarray}
F(x_n) &=& 
\gamma \vs + \sqrt{2\gamma T} \sum_{k=1}^\infty K_0 (t_k) \Delta W_{n-k} 
\nonumber \\
& & + \sqrt{2 \gamma T} 
\sum_{k =1}^\infty K_\perp (t_k, x_{n-k}) \Delta W_{n-k},
\label{e.p4}
\end{eqnarray}
it follows from Eq.~(\ref{e.Kp0})  that 
\begin{equation}
\bra \bar F_n \Delta W_n \ket_0 = \sqrt{\frac{\gamma T}{2}} K_0(\Delta t) \Delta t,
\label{e.p5}
\end{equation}
In addition, using Eq.~(\ref{e.rk0}), Fourier's integration theorem,
$
\lim_{\Delta t \to 0} [K_0(-\Delta t)+K_0(\Delta t)]/2
=\int_{-\infty}^\infty
\tilde K'_0 (\omega) \dw/2\pi,
$
leads  to 
\begin{equation}
\lim_{\Delta t \to 0+} K_0(\Delta t)
= 2\int_{-\infty}^\infty 
\left [\gamma \tilde R'(\omega)-1\right]\frac{\dw}{2\pi}.
\label{e.rint}
\end{equation}
Substituting Eqs.~(\ref{e.cint}), (\ref{e.p5}) and (\ref{e.rint}) 
into Eq.~(\ref{e.p3}), we obtain the theorem.


We now present several discussions on the generality of the equality 
proved here. First, although this proof has been restricted to Model B, 
we can prove Eq.~(\ref{e.equality}) in almost the same manner for Model A, 
by determining the decomposition of the force as expressed in 
Eq.~(\ref{e.decomp}). Furthermore, our result represented in 
Eq.~(\ref{e.equality}) can be generalized for a larger class of Brownian
ratchet models \cite{Reimann}, including models with an inertia effect, a time-periodic
potential, and  spatially inhomogeneous temperature profile. 
We will provide  a detailed description of these derivations in another paper.

Second, we can also analyze systems with many degrees of freedom, 
such as colloidal dispersions, in a parallel way.
As a simple example,  we consider a system where $N$ spherical 
colloidal particles in a three dimensional aqueous solution are 
driven by an constant external force $f \Vec{e}_x$. (See  Ref.~\cite{gier}
for a related experimental system.) Let us denote 
the coordinates of the particles by $\Gamma = \{ x_i \}$ ($ i = 1, \cdots, 3N$),
where $\Vec{r}_\mu \equiv (x_{3\mu-2} ,x_{3\mu-1}, x_{3\mu})$ represents the position of the $\mu$-th particle ($\mu = 1, \cdots, N$). Then, a widely used model describing the motion of the particles is provided as \cite{c-model}
\begin{eqnarray}
\gamma \dot x_i (t) = F_i (\Gamma (t)) + \xi_i (t) + \ve \fp_i (t),
\label{e.many}
\end{eqnarray}
where $F_i (\Gamma) = \sum_{\mu=1}^N f \delta_{i, 3\mu-2} - \partial_{x_i} \sum_{\mu=1}^N U(\Vec{r}_\mu) - \partial_{x_i} \sum_{\mu, \nu=1}^N 
U_{\mu\nu}^\mathrm{int} (|\Vec{r}_\mu - \Vec{r}_\nu|)/2$ represents single-body 
forces and two-body interactions, and the 
noise satisfies that 
$\bra \xi_i (t) \xi_j (s) \ket_0 = 2 \gamma T\delta_{ij} 
\delta(t-s)$.
For this model, the energy dissipated into 
the solvent is expressed as 
$J(t) \dt \equiv \sum_{i=1}^{3N} \left[ \gamma \dot x_i (t) 
- \xi_i (t) \right] \circ \dx_i(t)$ \cite{footnote}.
We can then  prove 
\begin{equation}
\bra J \ket_0 = \sum_{i=1}^{3N} \gamma \left\{ \bra \dot x_i \ket_0^2 + \int_{-\infty}^\infty \left[ \tilde C_{ii} (\omega) - 2T \tilde R'_{ii} (\omega) \right] \frac{\dw}{2\pi} \right\},
\label{e.manyeq}
\end{equation}
where $C_{ij} (t) \equiv \bra [\dot x_i(t) - \bra \dot x_i \ket_0]
[\dot x_j(0) - \bra \dot x_j \ket_0 ] \ket_0$ are the cross 
correlations of velocity fluctuations, and $R_{ij} (t)$ are 
the cross response functions defined as
\begin{equation}
\bra \dot x_i (t) \ket_\ve = \bra \dot x_i \ket_0 + \ve \sum_{j = 1}^{3N} \int_{-\infty}^t R_{ij} (t-s) \fp_j (s) \ds + O(\ve^2).
\end{equation}
The derivation of Eq.~(\ref{e.manyeq}) will be presented 
in another paper, although it is a straightforward extension of 
the above argument for Model B.

In addition, we present another generalization of Eq.~(\ref{e.equality}).
For Model B, by a similar argument with the above proof, we can derive an expression for the symmetrized time correlation 
between the velocity and force $ I(t) \equiv \bra [\dot x(t) \circ F(x(0)) +\dot  x(0) \circ F(x(t))]\ket_0 /2 $ as 
\begin{equation}
I(t) = 
\gamma \left\{ \vs^2 +\int_{-\infty}^\infty 
\left[ \tilde C(\omega) - 2T R'(\omega) \right] 
e^{i \omega t} \frac{\dw}{2\pi} \right\}.
\label{e.gen}
\end{equation}
In case of equilibrium, this equality leads to the FRR [Eq.~(\ref{e.FDT})] since we can derive $I(t)=0$ and $\vs = 0$ from the detailed balance condition. 
Thus, in order to further investigate the physical meanings of the FRR violation, it might be important to study the function $I(t)$.


Before we conclude, let us discuss the physical significance of the theorem.
From an experimental point of view, it has been difficult to estimate the 
amount of the rate of energy dissipation in systems under investigation.
The virtue of the expression in Eq.~(\ref{e.equality}) is that 
it enables us to determine the energy dissipation from 
experimentally accessible quantities alone, without knowing the every detail of the system such as the profile of the force, $F(x,t)$.
It is expected that the present result is also useful in the study of biological molecular motors.


In conclusion, we have presented an equality between the rate of energy dissipation and the extent of the FRR violation for a class of Langevin systems under non-equilibrium conditions.
To the best of our knowledge, no previous study has addressed this equality, with the exception of a similar, but not precise, expression that was conjectured  by one of the authors \cite{harada2}.
However, we should state that an inequality between the rate of entropy production and the FRR violation was demonstrated in Ref.~\cite{CDK}.

In the present Letter, we restricted our argument to Langevin dynamics.
To extend our result for more general non-equilibrium systems that are not
described by Langevin dynamics is an important open problem. Related 
to this problem, one might be able to prove the present result on 
the basis of microscopic dynamics, e.g., the Liouville equation, 
at least with focusing on the vicinity of equilibrium.
Experimental examinations on 
various non-equilibrium systems are also of great importance.

\begin{acknowledgments}
We  acknowledge K. Hayashi for the constant discussion with regard to all the issues in this Letter.
This work was supported by a grant from the Ministry of Education, Science, Sports and Culture of Japan, No. 16540337, and Research Fellowships for Young Scientists from the Japan Society for the Promotion of Science, No. 05494.
\end{acknowledgments}


\begin{thebibliography}{99}

\bibitem{ft}
D. J. Evans, E. G. D. Cohen, and G. P. Morriss, Phys. Rev. Lett. {\bf 71}, 2401 (1993).

\bibitem{cg}
G. Gallavotti and E. G. D. Cohen, Phys. Rev. Lett. {\bf 74}, 2694 (1995).

\bibitem{exft}
G. M. Wang et al., Phys. Rev. Lett. {\bf 89}, 050601 (2002).

\bibitem{jarz}
C. Jarzynski, Phys. Rev. Lett. {\bf 78}, 2690 (1997).

\bibitem{exjarz}
F. Ritort, C. Bustamante, and I. Tinoca, Jr., Proc. Natl. Acad. Sci. USA {\bf 99}, 13544 (2002).

\bibitem{hata}
T. Hatano and S. -i. Sasa, Phys. Rev. Lett. {\bf 86}, 3463 (2001).

\bibitem{exhata}
E. H. Trepagnier et al., Proc. Natl. Acad. Sci. USA {\bf 101}, 15038 (2004).

\bibitem{FDT}
R. Kubo, M. Toda, and N. Hashitsume, \textit{Statistical Physics II: Nonequilibrium Statistical Mechanics} (Springer-Verlag, Berlin, 1991).

\bibitem{CKP}
L. F. Cugliandolo, J. Kurchan, and L. Peliti, Phys. Rev. E {\bf 55}, 3898 (1997).

\bibitem{harada1}
T. Harada and K. Yoshikawa, Phys. Rev. E {\bf 69}, 031113 (2004).

\bibitem{HS1}
K. Hayashi and S. -i. Sasa, Phys. Rev. E {\bf 69}, 066119 (2004).


\bibitem{flash1}
R. D. Astumian and M. Bier, Phys. Rev. Lett. {\bf 72}, 1766 (1994).

\bibitem{flash2}
J. Prost, J. F. Chauwin, L. Peliti, and A. Ajdari, Phys. Rev. Lett. {\bf 72}, 2652 (1994).

\bibitem{Risken}
H. Risken, {\it The Fokker-Planck Equation} (Springer-Verlag, Berlin, 1996).

\bibitem{Sekimoto}
K. Sekimoto, J. Phys. Soc. Jpn. {\bf 66}, 1234 (1997).

\bibitem{gardiner}
C. W. Gardiner, {\it Handbook of Stochastic Methods for Physics, Chemistry and the Natural Sciences} (Springer-Verlag, Berlin, 2004).

\bibitem{2nd}
K. Sekimoto and S. -i. Sasa, J. Phys. Soc. Jpn. \textbf{66}, 3326 (1997).

\bibitem{Reimann}
P. Reimann, Phys. Rep. {\bf 361}, 57 (2002).

\bibitem{Parrondo}
J. M. R. Parrondo, J. M. Blanco, F. J. Cao and R. Brito, Europhys. Lett. {\bf 43}, 248 (1998).

\bibitem{HHS}
T. Harada, K. Hayashi, and S. -i. Sasa, J. Phys. A: Math. Gen. {\bf 38}, 3799 (2005).

\bibitem{gier} 
P. T. Korda, M. B. Taylor, and D. G. Grier, Phys. Rev. Lett. {\bf 89}  
128301, (2002).  

\bibitem{c-model} J. K. G. Dhont, {\it An introduction to dynamics of 
collois}, (Elsevier Science, Amsterdam, 1996).

\bibitem{footnote} This definition is also consistent with the linear response theory \cite{FDT}; it is easily shown that $\bra J \ket_0 = \sum_{\mu, \nu} \tilde R^{(f=0)}_{3\mu -2,3\nu-2}(0) f^2 + O(f^3)$ for sufficiently small $f$, where $R^{(f=0)}_{ij}(t)$ are the response functions in the $f \to 0$ limit. This expression of $\bra J \ket_0$ should be equivalent to Eq.~(\ref{e.manyeq}) in the small $f$ limit.

\bibitem{harada2}
T. Harada, Europhys. Lett. {\bf 70}, 49 (2005).

\bibitem{CDK}
L. F. Cugliandolo, D. S. Dean, and J. Kurchan, Phys. Rev. Lett. {\bf 79}, 2168 (1997).

\end{thebibliography}
\end{document}